# Evidence of Silicon Band-Edge Emission Enhancement When Interfaced with SiO$_2$:Er Films[†]

S. ABEDRABBO,[1,2,4] A.T. FIORY,[3] and N.M. RAVINDRA[3,5]

1. – Physics Department, CAS, The Petroleum Institute, Abu Dhabi, 51133 UAE. 2. – Department of Physics, The University of Jordan, Amman 11942 Jordan. 3. – Interdisciplinary Program in Materials Science & Engineering, New Jersey Institute of Technology, Newark, NJ 07102 USA. 4. – e-mail: sxa0215@yahoo.com. 5. – e-mail: nmravindra@gmail.com.

Nearly two-orders of magnitude increase in room-temperature band-to-band (1.067 eV) infrared emission from crystalline silicon, coated with erbium-doped sol-gel films, have been achieved. Phonon-assisted band-to-band emission from coated and annealed p-Si is strongest for the sample annealed at 700ºC. In this paper, evidence of the origin of the emission band from the band edge recombination activities is established. Enhancement of radiative recombination of free carriers is reasoned by stresses at the interface due to the annealed sol-gel deposited silica. Comparative studies with other strained silicon samples are presented.

(14 December 2016)

## INTRODUCTION

Fabricating efficient light emitting all-silicon devices has proven to be elusive despite the tremendous research efforts and the enormously painstaking capital expenditure thus far. This fact is predominantly due to the very nature of the indirect band-gap of silicon that circumvents efficient radiative recombination of free carriers (electron-hole) because of its continuous requirement of a third body (phonon) of equal and opposite wave-vector to the electron to secure conservation of momentum. The involvement of electron-hole-phonon makes this process an inefficient one [1]; typical quantum efficiency of $10^{-4}$ have been achieved at 300 K [2, 3]. In order to alleviate this hurdle, various attempts to fabricate engineered structures include the following: nanostructured $p$–$n$ junctions [4], oxidized porous Si [5, 6], Si nanocrystals embedded in suboxide [7], ultra-thin Si crystal (silicon on insulator or SOI) [8, 9], nanopatterned SOI, including enrichment with carbon [10], and $p$–$n$ junctions formed by ion implantation [11-15], diffusion doping [14] or amorphous-Si heterostructures [16]. These materials and structures provide a variety of approaches including band-structure engineering, quantum confinement in low-dimensional structures, carrier confinement, and impurity- or defect-mediated emission. The various methods relax competition with Auger recombination processes, enhance overlap of the electron and hole wave-functions, or reduce phonon involvement by band-alignment engineering.

In this work, novel silicon-based optical energy generating structures and functions, created by interfacing sol-gel films with crystalline silicon surfaces, are investigated. Substantially enhanced emission, of approximately 100 times, of infrared radiation near band-edge, i.e. 1.067 eV or 1162 nm is achieved [17]. The following sections will provide a proof that band-gap originated recombination activities are the source for the observed photoluminescence and stresses and strains are hypothesized as the mechanism for emission enhancement.

## EXPERIMENTAL

Samples for this study have been processed by spin-coating thin films of silica doped with erbium that is subsequently baked and cured by annealing at various temperatures. The sol–gel is prepared by adding 0.5 g $Er_2O_3$ powder to a solution of 4 mL ethanol, 4 mL acetic acid and 1.6 mL deionized water and stirred at 45 °C for 3 h. The source of silica is then presented by adding 2 mL of tetraethyl orthosilicate (TEOS) and further stirring for 10 min at 80 °C. This gel is then applied to a 1200 rpm spinning, 150 mm, p-type, RCA cleaned, double-side-polished, Czochralski (CZ) Si wafer via syringe filter with 0.45 $\mu$m pore size. The resulting film is baked in air at 120 °C for 30 min and contains 6 at. % Er. The processed wafer is then cut into small squared samples of side 2.5 cm that is annealed for 1 h under vacuum of 2 Pa, at temperatures ranging from 500-900ºC to cure the films, disperse the OH contaminants and achieve stressed interface with the underneath Si-substrate [18] but not to start a major thermal reflow that results in homogeneity in engineered stresses and strains [19].

Photoluminescence (PL) spectra are recorded by a model Fluorolog-3 spectro-fluorometer (Horiba Jobin Yvon) that utilizes double-excitation monochromator (Xe lamp source), single-emission monochromator, and cooled InGaAs photodiode (Hamamatsu), in the wavelength range of 850–1600 nm and excitation wavelength (522 nm).

## RESULTS AND DISCUSSION

### Light Emission from Si Band-Edge

One of the instigations during the few initial studies, that started as early as 2009, was to yield a low-cost and efficient erbium doped waveguide via sol-gel techniques. Annealing the samples in the range of 500-900ºC achieves the following: (1) Er-incorporation in the silica-glassy network, (2) optical activation of Er and (3) exploration of the nature of effect of stresses and strains, known to be sensitive to annealing temperature, on the optical activity of the Si-substrate.

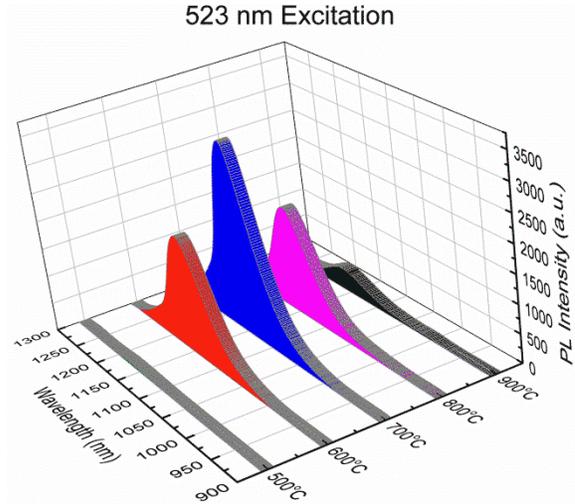

Fig. 1. 523-nm excited room-temperature photoluminescence spectra of $SiO_2$:Er/Si sample annealed at 500°C, 600°C, 700°C, 800°C and 900°C.

This entails that the Er bonds to $SiO_2$ and does not form segregated inclusions. Figure 1 exhibits the PL spectra of samples annealed at 500ºC, 600ºC, 700ºC, 800ºC and 900ºC in the wavelength range of 850–1300 nm. While objectives (1) and (2) have been achieved in the literature [17, 18], the authors excluded the well established 1500-nm Er-emission-band as it falls outside the scope of this work. Figure 1 shows the emission band exhibited at ~1162 nm (~1.067 eV) and is indicative of the band-edge radiative recom-bination activities along with weaker emissions in the region from 1.25–1.35 eV that are attributed to $^4I_{11/2} \rightarrow {}^4I_{15/2}$ transitions in $Er^{3+}$ which are obscured in this 3D plot. The 1.067 eV emission was investigated earlier [17] and has since then stemmed many more studies and new analysis in search of the relationship of processing methodology and annealing temperatures with the band-edge emission. This research explores some of these efforts and offers a better understanding of the probable cause for the unexpected Si-related emission with further evidence that the luminescence is indeed from Si.

As can be seen from the figure, the best sample for the highest emission is the one annealed at 700°C, followed by the one annealed at 600°C, 800°C, 900°C, while the sample annealed at 500°C exhibited miniscule emissions. Samples with thermally untreated coatings did not exhibit any luminescence and so did the uncoated p-Si; hence, one can consider the sample annealed at 500°C as the control sample for comparison, with the reported twofold emission enhancement. The thickness of the Er-doped silica coating in the 700°C annealed sample is 0.13 μm with shrinkage of ~25% from the un-annealed coated sample [17]. In Ref. 17, Abedrabbo et al. have reported a simple model for the emission of photons at energy $E$ by phonon-assisted recombination of free carriers in Si, thus showing that the emission band is real and is originating from the band-edge of crystalline Si. Such unexpected radiative recombination of un-patterned p-Si must stem from the band-gap perturbation of some sort that discourages the known indirect band-gap caveats. The only obvious source of such perturbation is in the stresses and strains formed by the interfacing of the top annealed spun-coated $SiO_2$:Er films in a manner that shield the readily available free-carriers from non-radiative recombination mechanisms. At 500°C, the silica is very porous and the glassy network that includes the densifying Er-element is barely forming and this translates into weaker stress fields that are not known to effectively perturb the band-gap of the underneath Si. The work of Parrill on spin-coated sol–gel films on Si has demonstrated that the 700°C stands out as the one yielding the broadest $TO_3$ modes known to be associated with the asymmetric Si–O–Si stretching vibrations [20]. This broadening is directly related to the strength of interfacial stresses and strains [21]. These findings triggered the need for a dedicated comparative study against stressed standard sample and hence the PL analysis in the next section.

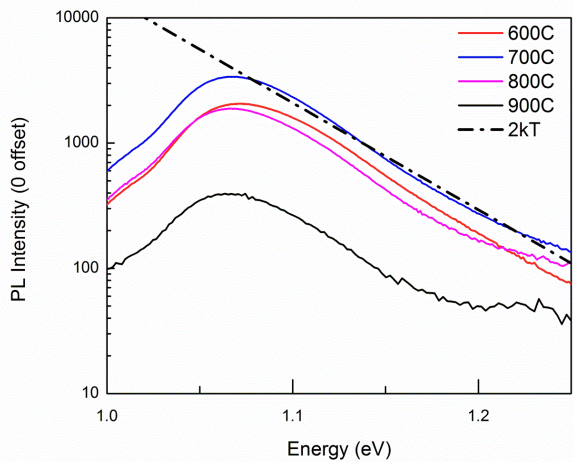

Fig. 2. Room-temperature short-wavelength part of the photoluminescence of the five samples plotted on a semi-logarithmic scale. The dashed-straight line is of slope $|2k_BT|$.

It has been known that the PL spectra are sensitive to the temperature of measurements and that, as the measurement temperature is increased, the temperature broadens the PL spectra. Figure 2 shows that the intensity of the high-energy tail of the PL spectrum for samples with appreciable emission at Si band-edge is linear on a semi-logarithmic scale and reasonably correlates for all the samples, within reasonable ~10% error, with line of slope of $\sim|2k_BT|$, where $k_B$ is the Boltzmann constant.

Although a bit oversimplified, this analysis of the high-energy tail of the PL spectra follows an observation reported previously by Tsybeskov et al. [5]. Recombination of free-carriers (electrons and holes) across the indirect band-gap of Si entails readily available electrons in the conduction band that are statistically predicted as function of temperature by the probability of occupied energy states or $f(E,T)$. Multiplying this probability by the density of states $N(E)$ yields the total broadening of the PL line, i.e. $\Delta E \sim f(E,T) \times N(E)$, that is proportional to $2k_BT$ [5]. Plotting the PL axis on a semi-log scale has captured the linear drop of the high-energy tail with the linear region 1.112 eV $< E <$ 1.268 eV fitting nicely (within reasonable error) to a linear line of slope of 0.05084 eV ($2k_BT$, and $T = 295$K), mainly for the samples annealed at 600ºC and 700ºC. Beyond 1.2 eV, various samples behave differently due to the intensity of the weak 980 nm band attributed to the $^4I_{11/2} \rightarrow {}^4I_{15/2}$ transitions in $Er^{3+}$, unrelated to Si, available at the interface in the silica films. It can be seen that, beyond 1.2 eV, the samples annealed at 800ºC and 900ºC exhibit higher deviations from the straight line because the Er-related optical activity is enhanced [22]. It may be noted that the intrinsic carrier density, $n_i \propto \exp(-\frac{E_g}{2kT})$, related to $2k_BT$, may indicate the involvement of an intrinsic behavior, perhaps as a limit, on the radiative recombination of the stressed interface that causes this behavior. This also corroborates with the uncorrelated pair model [22] and its manifestation in reference [17] in pointing out the source of the unexpected optical activity – namely, the Si band edge.

**Photoluminescence analysis**

The photoluminescence spectra of $SiO_2$:Er/Si samples include the combined emissions from $Er^{+3}$ centers in the $SiO_2$:Er film and phonon-assisted electron-hole recombination in the Si substrate. Since the two contributions originate from different material sources, the Si emission is analyzed separately from the Er emission. The 600°C-annealed sample is selected for analysis, since Si yields relatively stronger emission than Er, when compared to the 700°C-annealed sample that was previously treated with an elementary one-phonon model [17].

Phonon-assisted emission from Si is modeled as a sum over the momentum-conserving phonon emission and absorption processes. The photoluminescence intensity PL($E$) from the emission of photons of energy $E = hc/\lambda$, where $\lambda$ is the emission photon wavelength, is modeled by the expression,

$$PL(E) = \sum_i A_i (E - E_G - E_i)^n \exp[-(E - E_G - E_i)/k_BT] \;,$$

(1)

where $E_i$ is the energy of one or two phonons involved in a given process denoted by index $i$; amplitudes $A_i$ are taken to be independent of $E$ and implicitly include phonon assisted thermal excitation probability factors. Equation 1 follows from Eq. 6 of Ref. 16 augmented with the phonon-related components from Ref. 23. $E_G$ is the silicon band gap and $T$ is the ambient temperature, taken to be 295 K in this analysis.

The exponent $n$ acquires the value $n = 2$ for uncorrelated electron-hole recombination, reflecting the parabolic band edge density of electronic states. For correlated recombination, one has $n = 0.5$, as in the case of Si electroluminescence, where a portion of correlated recombination is observed at room temperature [16]. The energies $E_i$ are written in terms of the transverse optical (TO) phonon and TO in combinations with intervalley phonons ($IV_1$ and $IV_2$), as shown in Table I (see Ref. 23), where index $i$ ranges from 1 to 6. Contributions to photoluminescence of $SiO_2$:Er/Si from transverse acoustic (TA) phonons are found to be essentially absent, and thus TA phonons are not listed in Table I.

Photoluminescence data for the $SiO_2$:Er/Si sample, annealed at 600°C, is shown in Fig. 3 (excitation at 523 nm), normalized to unity at maximum $\lambda = 1157.4$ nm. The peak in Fig. 3 is formed mainly by the phonon-assisted emissions from the Si substrate material. Two changes in slope are evident at $\lambda \approx 1220$ nm and $\lambda \approx 1294$ nm, which correspond to emission thresholds at the Si band edge where $E \approx E_G + E_i$ for $i = 1$ and 4, respectively. The small emission from the $Er^{+3}$ in the $SiO_2$:Er film, occurring in the vicinity of 1500 nm, and the minor component near 925 nm are not included in the model of Eq. 1 and Table I.

**Table I.** (a) Phonon modes and energies; (b) phonon processes and energies $E_i$ for phonon creation (+) and annihilation (−)

(a)

| Phonon mode | Phonon energy (eV) |
|---|---|
| TO | 0.058 |
| $IV_1$ | 0.060 |
| $IV_2$ | 0.023 |

(b)

| $i$ | Process | $E_i$ (eV) |
|---|---|---|
| 1 | +TO | 0.058 |
| 2 | −TO | −0.058 |
| 3 | +TO + $IV_1$ | 0.118 |
| 4 | +TO + $IV_2$ | 0.081 |
| 5 | +TO + 2 $IV_1$ | 0.178 |
| 6 | +TO + $IV_1$ + $IV_2$ | 0.141 |

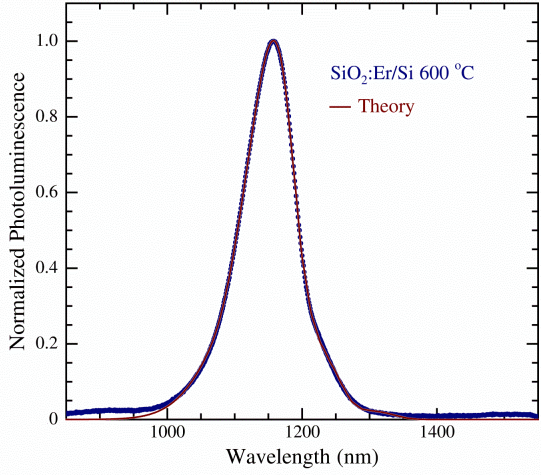

Fig. 3. Photoluminescence spectrum of $SiO_2$:Er/Si sample annealed at 600°C (blue points) and theoretical model function (red curve).

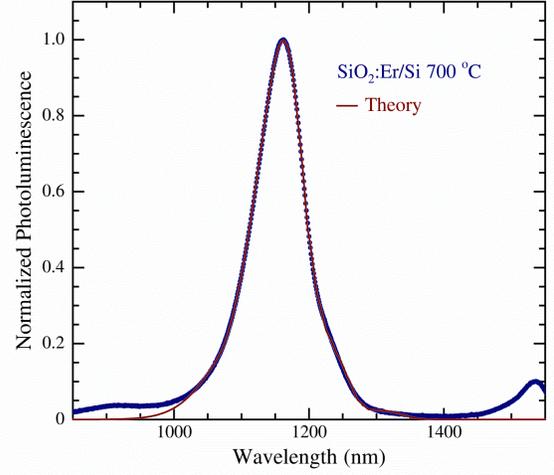

Fig. 4. Photoluminescence spectrum of $SiO_2$:Er / Si sample annealed at 700°C (blue points) and theoretical model function (red curve).

The spectrum of Fig. 3 has been modeled according to Eq. 1 using the phonon parameters for $E_i$ in Table I. Observing that the above mentioned changes in slopes in Fig. 3 appear non-abrupt and somewhat smeared out, the function of Eq. 1 is convolved with a Gaussian function by applying a 7-point Gauss-Hermite quadrature digital integration [24], to obtain the fitting function as,

$$PL_{FIT}(E) = \sum_{j=1}^{7} w_j PL(E + \sqrt{2}\sigma x_j), \qquad (2)$$

where, $w_j$ and $x_j$ are the quadrature weights and roots, respectively, and $\sigma$ is the fitted Gaussian broadening parameter. A non-linear least squares fitting routine was used to obtain the best fit of $PL_{FIT}(E)$ to the photoluminescence spectrum by varying the parameters $A_i$, $E_G$, $n$, and $\sigma$. The resulting theoretical function is shown by the red curve in Fig. 3. In the region generating most of the Si emission (990 nm < $\lambda$ < 1370 nm), the fitting error is 0.4% rms of full scale. The fitted parameters are given in the second column of Table II, where $A_i$ is normalized to $A_1$ for the TO-dominant component. Equation 2 similarly, captures the Si emission from the sample annealed at 700°C, albeit with greater systematic parameter uncertainties from overlapping Er emission.

The photoluminescence spectra for the $SiO_2$:Er / Si sample annealed at 700°C is shown in Fig. 4, where the red curve is the theoretical function fitted with the parameters given in third column of Table II.

For making a comparison to a sample with an optically inactive film, the same theoretical analysis was applied to the room-temperature photoluminescence spectrum (650 nm excitation) obtained for a SiN/$SiO_2$/Si sample with 2.2 nm $SiO_2$ grown at 750 °C in an oxidation furnace and 5.7 nm SiN deposited by atomic layer deposition in ammonia at 620°C (from Fig. 2a of Ref. 25, sample F). The utilized process yields a compressive stress of about 20 MPa in the Si [25]. The Raman shift relative to stress-free Si was determined to be $\Delta\omega = +0.14$ cm$^{-1}$ at an excitation wavelength of 457 nm (~290 nm probe depth) [25]. The (001)-plane biaxial compressive strain is estimated from the expression $\varepsilon_\parallel = c\Delta\omega$ using the experimentally determined value of $c \approx 1.33 \times 10^{-3}$ cm [26], yielding $\varepsilon_\parallel \approx 2\times 10^{-4}$. Hence, the SiN/$SiO_2$/Si sample provides a relatively low-strain reference.

**Table II.** Fitted parameters modeling PL(*E*) for SiO$_2$:Er/Si samples annealed at 600°C and 700°C and a SiN/SiO$_2$/Si reference sample [25]; statistical uncertainties in least significant digits are given in parentheses

| Parameter | SiO$_2$:Er/Si 600°C | SiO$_2$:Er/Si 700°C | SiN/SiO$_2$/Si |
|---|---|---|---|
| $A_1$ | 1.00(2) | 1.00(2) | 1.00(1) |
| $A_2$ | 0.038(1) | 0.051(1) | 0 |
| $A_3$ | 0.030(1) | 0.250(6) | 0.101(3) |
| $A_4$ | 0.296(18) | 0.366(13) | 0.105(4) |
| $A_5$ | 0.030(1) | 0.034(1) | 0.016(1) |
| $A_6$ | 0.007(3) | 0.0(1) | 0.006(2) |
| $E_G$ (eV) | 1.0907(1) | 1.0924(2) | 1.103(1) |
| $n$ | 1.62(1) | 1.38(8) | 1.74(1) |
| $\sigma$ (eV) | 0.0106(3) | 0.0150(3) | 0 |

The photoluminescence spectra (green curve) and the theoretical fit to the SiN/SiO$_2$/Si data (red curve) are shown in Fig. 5. The fitted parameters are given in the fourth column of Table II.

For both samples, $E_G$ is smaller than that for intrinsic Si, 1.11 eV. The values of $E_G$ in Table II are uncorrected for possible exciton formation, which, if present, increases $E_G$ by the exciton binding energy of 0.015 eV.

**Discussion**

When compared to the results for the relatively low stress SiN/SiO$_2$/Si sample, taken as a reference, several distinguishing characteristics of the SiO$_2$:Er/Si sample are worth noting.

The value of $E_G$ = 1.0907(1) eV for SiO$_2$:Er/Si indicates band-gap narrowing by 0.012(1) eV relative to SiN/SiO$_2$/Si. In biaxially tensile strained Si layers, shifts in the TO-dominant photoluminescence peaks have been interpreted in terms of strain-induced narrowing of the Si band-gap [26]. Using the results of such measurements, the energy band-gap deformation potential is estimated as $a_\parallel$ =

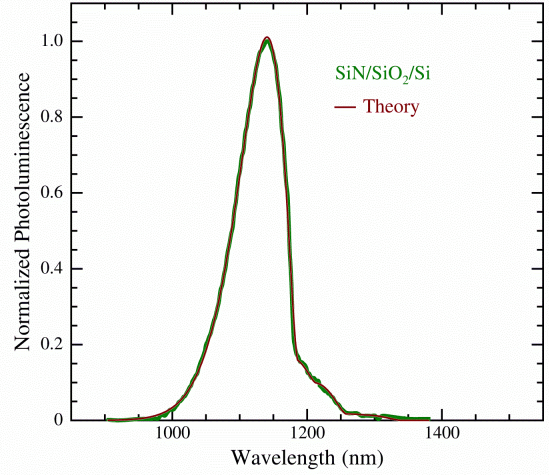

Fig. 5. Photoluminescence spectrum of a SiN/SiO$_2$/Si sample from Ref [25] (green curve) and theoretical model (red curve).

$\Delta E_{TO}/\varepsilon_\parallel \approx 11$ eV. For the SiO$_2$:Er/Si sample, one thus estimates $\varepsilon_\parallel \approx 0.012$ eV / $a_\parallel$ − 0.0002 = 0.0009, wherein the finite strain in the reference SiN/SiO$_2$/Si is subtracted.

The photoluminescence spectrum for SiO$_2$:Er/Si is broadened by $\sigma$ = 0.0106 eV. In contrast, photoluminescence from the SiO$_2$:Er/Si sample is essentially unbroadened; the spectrum from porous Si exhibited in Ref. 5 looks similarly unbroadened. From the broadening parameter $\sigma$, the inhomogeneous component of the strain in SiO$_2$:Er/Si is estimated as $\delta\varepsilon_\parallel = \sigma/a_\parallel$ = 0.001. Correction for the instrumental width is deemed negligible, given the 0.5-nm resolution of the spectrophotometer.

The phonon absorption process contributes 3.8(1)% to TO phonon activity, whereas TO absorption is negligible in the reference sample. Differences in the TO absorption component $A_2$ may reflect experimental differences in optical excitation power for the spectra shown in Figs. 3, 4 and 5. Although the Si and Er emissions are found to be uncorrelated [17], the roles of Er concentration and film thickness remain to be determined in future work.

## CONCLUSION

Interfacing ordinary and unpatterned crystalline Czochralski p-Si with spun-coated and annealed SiO:Er films greatly enhanced its luminescence efficiency. The involved and unexpected increase in emission efficiency in the indirect band-gap semiconductor is the result of interfacial stresses and strains that are believed to perturb the band-gap and to shield the free-carriers from known non-radiative recombination mechanisms.

Comparative studies with ordinary p-Si references of known levels of stresses by totally different processes that exhibited enhanced optical emission efficiency at room-temperature are performed. Useful information about the FWHM, band gap energy, involvement of various types of phonons has been uncovered. Noteworthy data of resemblances on FWHM of the two samples of unrelated processes revealed evidence that the band-edge is the source for the enhanced optical activity.

## ACKNOWLEDGEMENTS

Partial support by the Deanship of Academic Research at the University of Jordan, Project Contract No. 1030 and Hamdi Mango Center for Scientific Research (HMCSR) is acknowledged with thanks. The authors would like to acknowledge Dr. Bashar Lahlouh and Dr. Sudhakar Shet for their support and interest.